\documentclass[aps,prl,superscriptaddress,showpacs,twocolumn,
groupedaddress,nofootinbib,nobalancelastpage,nobibnotes]{revtex4}
\usepackage{amsmath,amsfonts,amssymb,mathrsfs,graphicx}

\newcommand{\ie}{{\it i.e.}}

\newcommand{\eq}{Eq.}

\newcommand{\fig}{Fig.}

\newcommand{\Ref}{Ref.}
\newcommand{\Refs}{Refs.}

\begin{document}

\vspace*{-1.45cm}
\begin{flushright}
TUM-HEP-499/03~~~MPI-PhT/2003-05
\end{flushright}

\title{Neutrino Factories and the ``Magic'' Baseline}

\author{Patrick Huber}
\email[E-mail address: ]{phuber@ph.tum.de}
\affiliation{Max-Planck-Institut f\"ur Physik, Postfach 401212,
       D--80805 M\"unchen, Germany}
\affiliation{Institut f{\"u}r Theoretische Physik, Physik-Department,
Technische Universit{\"a}t M{\"u}nchen (TUM), James-Franck-Stra\ss{}e,
85748 Garching bei M{\"u}nchen, Germany}

\author{Walter Winter}
\email[E-mail address: ]{wwinter@ph.tum.de}
\affiliation{Institut f{\"u}r Theoretische Physik, Physik-Department,
Technische Universit{\"a}t M{\"u}nchen (TUM), James-Franck-Stra\ss{}e,
85748 Garching bei M{\"u}nchen, Germany}

\date{\today}

\begin{abstract}
\vspace*{0.2cm}
We show that for a neutrino factory baseline of $L \sim 7 \, 300 \,\mathrm{km} - 7 \, 600 \, \mathrm{km}$ a ``clean'' measurement of $\sin^2 2 \theta_{13}$ becomes possible, which is almost unaffected by parameter degeneracies. We call this baseline ``magic'' baseline, because its length only depends on the matter density profile. For a complete analysis, we demonstrate that the combination of the magic baseline with a baseline of $3 \, 000 \, \mathrm{km}$ is the ideal solution to perform equally well for the $\sin^2 2 \theta_{13}$, sign of $\Delta m_{31}^2$, and CP violation sensitivities. Especially, this combination can very successfully resolve parameter degeneracies even below $\sin^2 2 \theta_{13} < 10^{-4}$.
\end{abstract}


\maketitle


In neutrino physics, there is now quite strong evidence for atmospheric and solar neutrino oscillations after the Super-Kamiokande and KamLAND results~\cite{Toshito:2001dk,unknown:2002dm}. In addition, the solar LMA (Large Mixing Angle) region is the only remaining region which can explain the solar neutrino oscillations~\cite{unknown:2002dm}. The coupling between the solar and atmospheric neutrino oscillations is described by $\sin^2 2 \theta_{13}$, which is bound by the CHOOZ-experiment to $\sin^2 2 \theta_{13} \lesssim 0.1$~\cite{Apollonio:1999ae}. The size of $\sin^2 2 \theta_{13}$ is, together with the solar $\Delta m_{21}^2$ lying within the LMA region, directly relevant for the detection of three-flavor effects in neutrino oscillations, such as leptonic CP violation. Three-flavor and other suppressed effects will be tested in future reactor and long-baseline experiments, such as superbeam and neutrino factory experiments (see, for a summary, \Ref~\cite{Apollonio:2002en}). Because of systematical errors, superbeams and superbeam upgrades are limited to $\sin^2 2 \theta_{13} \gtrsim 10^{-3}$, whereas neutrino factories can, in principle, be sensitive to three-flavor effects even below $\sin^2 2 \theta_{13} \lesssim 10^{-4}$. Thus, neutrino factories are a promising goal in the long-baseline roadmap. However, it has been demonstrated that neutrino factory measurements are spoilt by the presence of degenerate, often disconnected solutions in the neutrino oscillation formulas~\cite{Huber:2002mx}. Those are the $(\delta, \theta_{13})$~\cite{Burguet-Castell:2001ez}, $\mathrm{sgn}(\Delta m_{31}^2)$~\cite{Minakata:2001qm}, and $(\theta_{23},\pi/2-\theta_{23})$~\cite{Fogli:1996pv} degeneracies, \ie, and overall ``eight-fold'' degeneracy~\cite{Barger:2001yr}. Since the current best-fit value for the atmospheric angle is $\theta_{23}=\pi/4$, the first two of these degeneracies together with multi-parameter correlations strongly limit future long-baseline experiments~\cite{Huber:2002mx}. Several options to resolve this problem have been proposed, such as the combination of neutrino factories with superbeam upgrades~\cite{Burguet-Castell:2002qx} and the possibility to detect $\nu_\tau$'s in order to have additional oscillation channels with complementary information~\cite{Donini:2002rm}. Since one neutrino factory naturally forms two baselines and detectors are relatively cheap compared to the accelerator complex, the combination of two baselines to resolve degeneracies seems to be a straightforward choice. In this work, we will show that the appropriate combination of two neutrino factory baselines, where one of those is the magic baseline $L_{\mathrm{magic}} \sim 7 \, 300 \, \mathrm{km} - 7 \, 600 \, \mathrm{km}$, can resolve the degeneracies very competitively, an option which even works for $\sin^2 2 \theta_{13}$ much below $10^{-3}$.


For long-baseline experiments, the appearance probability $\nu_{e} \rightarrow \nu_{\mu}$ in matter can be expanded in the small hierarchy parameter $\alpha \equiv \Delta m_{21}^2/\Delta m_{31}^2$ and the small $\sin 2 \theta_{13}$ up to the second order as~\cite{CERVERA,FREUND}:
\begin{eqnarray}
P_{e\mu} & \simeq & \sin^2 2\theta_{13} \, \sin^2 \theta_{23} \frac{\sin^2[(1- \hat{A}){\Delta}]}{(1-\hat{A})^2} 
\nonumber \\
&\pm&   \alpha  \sin 2\theta_{13} \,  \xi \sin \delta_{\mathrm{CP}}   
\sin({\Delta})  \frac{\sin(\hat{A}{\Delta})}{\hat{A}}  \frac{\sin[(1-\hat{A}){\Delta}]}{(1-\hat{A})}
\nonumber  \\
&+&   \alpha  \sin 2\theta_{13} \,  \xi \cos \delta_{\mathrm{CP}} \cos({\Delta})  \frac{\sin(\hat{A}{\Delta})}{\hat{A}}  \frac{\sin[(1-\hat{A}){\Delta}]} {(1-\hat{A})}
 \nonumber  \\
&+&  \alpha^2 \, \cos^2 \theta_{23}  \sin^2 2\theta_{12} \frac{\sin^2(\hat{A}{\Delta})}{\hat{A}^2}.
\label{equ:PROBMATTER}
\end{eqnarray}
Here $\Delta \equiv \Delta m_{31}^2 L/(4 E)$, $\xi \equiv \cos\theta_{13} \, \sin 2\theta_{12} \, \sin 2\theta_{23}$,  and $\hat{A} \equiv \pm (2 \sqrt{2} G_F n_e E)/\Delta m_{31}^2$ with $G_F$ the Fermi coupling constant and $n_e$ the electron density in matter. The sign of the second term is determined by choosing $\nu_e \rightarrow \nu_\mu$ (positive) or $\nu_\mu \rightarrow \nu_e$ (negative) as the oscillation channel, and the sign of $\hat{A}$ is determined by the sign of $\Delta m_{31}^2$ and choosing neutrinos or antineutrinos. This formula shows that close to the resonance condition $\hat{A} \simeq 1$ especially the first term can be enhanced by matter effects. Therefore, it is most affected by the sign of $\Delta m_{31}^2$. In addition, CP effects are only present in the second and third terms. Depending on which quantity should be measured, one or two of the terms will act as signal and the rest of the terms as background. The formula also indicates that the values of $\sin 2 \theta_{13}$ and the hierarchy parameter $\alpha$ change the relative weight of the individual terms, which means that in certain regions of the $\sin^2 2 \theta_{13}$-$\alpha$-plane the measurements corresponding to the selected terms will be favored. For example, for CP violation measurements both $\alpha$ and $\sin 2 \theta_{13}$ should be large, and for the sign of $\Delta m_{31}^2$ and $\sin^2 2 \theta_{13}$ measurements $\alpha$ should be small. 

Many of the degeneracy problems originate in the summation of the four terms especially for large $\alpha$, since changing one parameter value in one term can be often compensated by adjusting another one in a different term. For instance, changing the sign of $\Delta m_{31}^2$ mostly affects the first term and can often be compensated by the second and third terms for a different value of $\delta_{\mathrm{CP}}$ (``$\mathrm{sgn} (\Delta m_{31}^2)$-degeneracy''). One strategy to circumvent this problem is choosing $\sin( \hat{A} \Delta) = 0$, which makes all but the first term in \eq~(\ref{equ:PROBMATTER}) disappear and thus allows a clean measurement of $\sin^2 2 \theta_{13}$ and the sign of $\Delta m_{31}^2$ without correlations with the CP phase~\cite{Lipari:1999wy,Barger:2001yr}.
 This condition is, for the first non-trivial solution, equivalent with $\sqrt{2} G_F n_e L = 2 \pi$, or, in terms of the constant matter density $\rho$, for approximately two electrons per nucleon, equivalent with 
\begin{equation}
 L_{\mathrm{magic}} \, [\mathrm{km}]  \simeq 32 \, 726 \, \, \frac{1}{\rho \, [\mathrm{g/cm^3}]}.
\end{equation}
Thus, it only depends on the matter density, but it does not depend on the energy and the oscillation parameters. Hence, we further on call this baseline the ``magic'' baseline~\cite{Huber:2002uy}. For a constant matter density, it evaluates to $L_{\mathrm{magic}} \simeq 7 \, 630 \, \mathrm{km}$ with the average matter density of this baseline $\rho \simeq 4.3 \, \mathrm{g/cm^3}$. Numerically, it can be shown to be closer to $L_{\mathrm{magic}} \sim 7 \, 250 \, \mathrm{km}$ for a realistic PREM (Preliminary Reference Earth Model) profile by minimizing the $\delta_{\mathrm{CP}}$-dependence in the appearance rates. 
For example, the baseline from Fermilab to Gran Sasso is magic in the sense of this definition.
Of course, the magic baseline has two obvious disadvantages: first, the statistics is pretty low at such a long baseline, and second, it does not allow a CP measurement, because the corresponding second and third terms in \eq~(\ref{equ:PROBMATTER}) are suppressed. We thus propose to combine the magic baseline with a shorter baseline with better statistics and the ability to access $\delta_{\mathrm{CP}}$. In this combination, the magic baseline would allow a clean measurement of $\sin^2 2 \theta_{13}$ and the sign of $\Delta m_{31}^2$, and, at the same time, with the shorter baseline a measurement of $\delta_{\mathrm{CP}}$ without much correlation with $\theta_{13}$. In the rest of this work, we demonstrate that this concept also works for a full statistical analysis.

\begin{figure}[t!]
\begin{center}
\includegraphics[width=8cm]{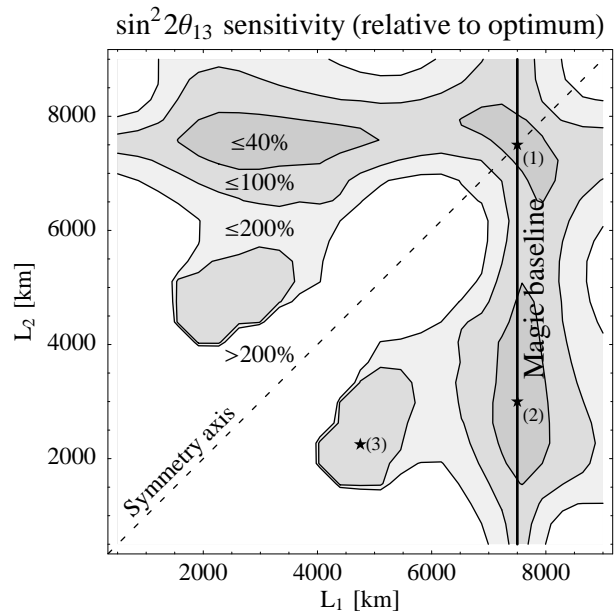}
\vspace*{-0.5cm}
\end{center}
\caption{\label{fig:master} The $\sin^2 2 \theta_{13}$ sensitivity limit relative to the optimum value of $5.9 \cdot 10^{-5}$ at $L_1=L_2 \simeq 7 \, 500 \, \mathrm{km}$. It is plotted at the $3 \sigma$ confidence level as function of the baselines $L_1$ and $L_2$ heading from the neutrino factory defined in the text towards two $25 \, \mathrm{kt}$-detectors. The sensitivity limits in this figure include systematics, multi-parameter correlations, and degeneracies and are computed for the LMA best-fit values as given in the text.}
\end{figure}

\begin{figure*}[t!]
\begin{center}
\includegraphics[width=17cm]{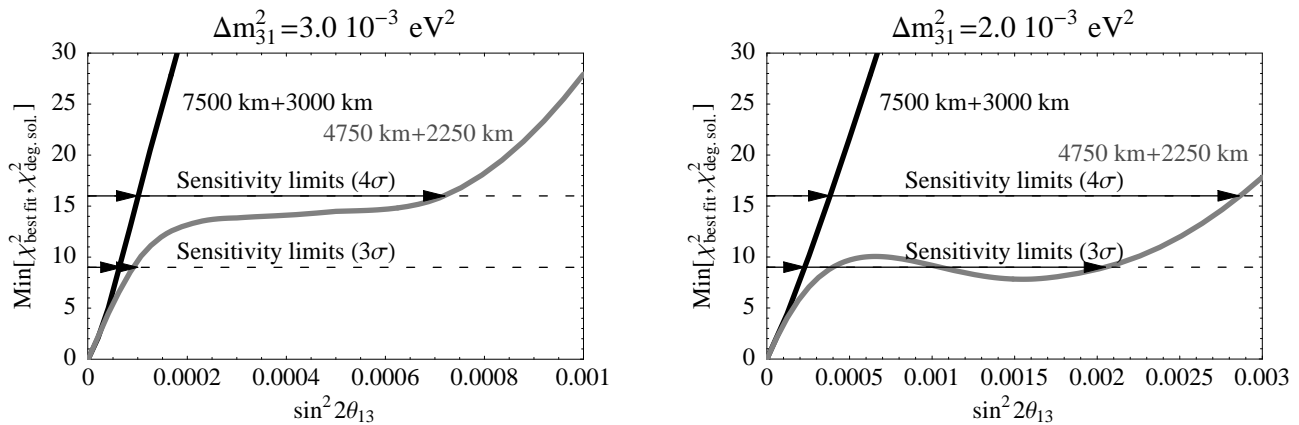}
\vspace*{-0.5cm}
\end{center}
\caption{\label{fig:chi2} The minimum of the $\chi^2$-values of the best-fit and degenerate regions as projections on $\sin^2 2 \theta_{13}$ for $\Delta m_{31}^2 = 3.0 \cdot 10^{-3} \, \mathrm{eV}^2$ (left plot) and $\Delta m_{31}^2 = 2.0 \cdot 10^{-3} \, \mathrm{eV}^2$ (right plot) for the true value $\sin^2 2 \theta_{13}=0$ and the neutrino factory described in the text. The black curves show this function for the combination of the $7 \, 500 \, \mathrm{km}$- and $3 \, 000 \, \mathrm{km}$-baselines and the gray curves for the combination of the $4 \, 750 \, \mathrm{km}$- and $2 \, 250 \, \mathrm{km}$-baselines. The arrows mark the sensitivity limits at the $3 \sigma$ and $4 \sigma$ confidence levels, respectively.}
\end{figure*}


For the analysis, we use the advanced stage neutrino factory scenario ``NuFact-II'' from \Ref~\cite{Huber:2002mx} with a muon energy of $50 \, \mathrm{GeV}$. It has a target power of $4 \, \mathrm{MW}$, corresponding to $5.3 \cdot 10^{20}$ useful muon decays per year. In addition, we assume eight years of total running time, four of these with a neutrino beam and four with an antineutrino beam. For the detector, we use a magnetized iron detector with an overall fiducial mass of $50 \, \mathrm{kt}$. However, since a neutrino factory naturally has two baselines, we split the detector mass into two equal pieces of $25 \, \mathrm{kt}$ each and place them at two baselines $L_1$ and $L_2$. The splitting into equal pieces can be justified by making the problem symmetric or using the same technology for the two detectors. We use the analysis technique which is described in the Appendices A, B, and C of \Ref~\cite{Huber:2002mx}, including the beam and detector simulations. For each baseline, we use a different average matter density corresponding to this baseline and allow an uncertainty of $5 \%$ on it. Furthermore, we assume the product $\Delta m_{21}^2 \cdot \sin 2  \theta_{12}$ of the solar parameters to be measured with $15\%$ precision by the KamLAND experiment by then~\cite{BARGER,Gonzalez-Garcia:2001zy}. We include systematics, multi-parameter correlations and the degeneracies in the analysis, as it is described in \Ref~\cite{Huber:2002mx} for the individual measurements. For the oscillation parameters, we choose the current atmospheric and solar (LMA-I) best-fit values $\Delta m_{31}^2 = +3.0 \cdot 10^{-3} \, \mathrm{eV}^2$, $\sin^2 2 \theta_{23} = 1.0$~\cite{Gonzalez-Garcia:2002mu}, $\Delta m_{21}^2 = 7.0 \cdot 10^{-5} \, \mathrm{eV}^2$, $\sin^2 2 \theta_{13} = 0.8$ (see, for example, \Ref~\cite{Maltoni:2002aw}), as well as we only allow values of $\sin^2 2 \theta_{13}$ below the CHOOZ bound. Furthermore, we do not make any special assumptions about $\delta_{\mathrm{CP}}$, \ie, we let it vary from $0$ to $2 \pi$ if not otherwise stated.


In order to establish the magic baseline, we need to demonstrate that the combination of some baseline with the magic baseline really is optimal in two-baseline-space $(L_1,L_2)$, since other combinations with shorter baselines could be better because of better statistics. Therefore, we show in \fig~\ref{fig:master} the $\sin^2 2 \theta_{13}$ sensitivity limit, \ie, the largest value of $\sin^2 2 \theta_{13}$ which cannot be distinguished from $\sin^2 2 \theta_{13}=0$, relative to its optimum in two-baseline space. In this figure, the symmetry axis corresponds to building both detectors at the same baseline, \ie, building one large detector instead of two smaller ones. Indeed, it demonstrates that all combinations of baselines, with one of them being magic, perform very well. The global optimum is in this figure at about $7 \, 500 \, \mathrm{km} + 7 \, 500 \, \mathrm{km}$~(1) and two local optima are located at about $7 \, 500 \, \mathrm{km} + 3 \, 000 \, \mathrm{km}$~(2) and $4 \, 750 \, \mathrm{km} +  2 \, 250 \, \mathrm{km}$~(3).  The first of these three solutions would not help us to measure CP violation since the magic baseline suppresses the CP-terms in \eq~(\ref{equ:PROBMATTER}). The second solution would certainly perform very well for CP violation, since the $3 \, 000 \, \mathrm{km}$-baseline is well-known to be good for this measurement~\cite{CERVERA}. The third solution is somewhat worse than the other two, but especially interesting since it lies off the magic baseline in the figure and it is especially favored by statistics because of the shorter baselines.
In \fig~\ref{fig:chi2}, we demonstrate the weakness of this solution: the $(\delta,\theta_{13})$-degeneracy may or may not lie below the chosen confidence level. For example, the degeneracy is present at the atmospheric best-fit value, but does not go under the $3 \sigma$-confidence level (left plot). However, it makes the sensitivity already worse at the $4 \sigma$ confidence level by almost an order of magnitude. In addition, it is not stable, such that it would affect the measurement for a worse energy resolution or energy threshold of the detectors~\cite{Huber:2002mx}, or different parameter values, by moving down under the chosen confidence level. This is illustrated in the right plot for a somewhat smaller value of $\Delta m_{31}^2$ within the Super-Kamiokande allowed region. Since the magic baseline does not suffer from this problem, such as it is shown in the plots for the $7 \, 500 \, \mathrm{km} + 3 \, 000 \, \mathrm{km}$ combination, we will therefore not discuss the $4 \, 750 \, \mathrm{km} +  2 \, 250 \, \mathrm{km}$ option anymore.


\begin{figure}
\begin{center}
\includegraphics[width=8cm]{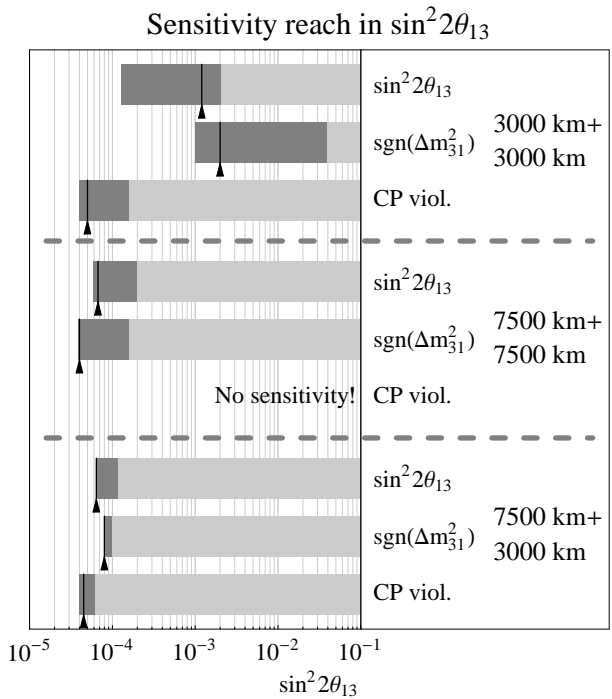}
\vspace*{-0.5cm}
\end{center}
\caption{\label{fig:sensreach} The sensitivity reaches as functions of $\sin^2 2 \theta_{13}$ for $\sin^2 2 \theta_{13}$ itself, the sign of $\Delta m_{31}^2>0$, and (maximal) CP violation $\delta_{\mathrm{CP}}=\pi/2$ for each of the indicated baseline-combinations. The bars show the ranges in $\sin^2 2 \theta_{13}$ where sensitivity to the corresponding quantity can be achieved at the $3 \sigma$ confidence level. The dark bars mark the variations in the sensitivity limits by allowing the true value of $\Delta m_{21}^2$ vary in the $3 \sigma$ LMA-allowed range given in \Ref~\cite{Maltoni:2002aw} and others ($\Delta m_{21}^2 \sim 4 \cdot 10^{-5} \, \mathrm{eV}^2 - 3 \cdot 10^{-4} \, \mathrm{eV}^2$). The arrows/lines correspond to the LMA best-fit value.}
\end{figure}

In addition to the $\sin^2 2 \theta_{13}$ sensitivity, there are several other measurements interesting for a neutrino factory. In order to demonstrate the potential of the magic baseline to resolve degeneracies, we choose two additional representatives: the sensitivity to a positive sign of $\Delta m_{31}^2$ and the sensitivity to maximal CP violation $\delta_{\mathrm{CP}}=\pi/2$, as they are defined in \Refs~\cite{Huber:2002mx,Huber:2002rs}. In \fig~\ref{fig:sensreach}, we show the corresponding sensitivity reaches for the following options: putting the whole detector mass to the $3 \, 000 \, \mathrm{km}$-baseline ($3 \, 000 \, \mathrm{km} + 3 \, 000 \, \mathrm{km}$) or the magic baseline ($7 \, 500 \, \mathrm{km} + 7 \, 500 \, \mathrm{km}$), or sharing the total detector mass between these two baselines in equal pieces ($7 \, 500 \, \mathrm{km} + 3 \, 000 \, \mathrm{km}$). In this figure, the dark bars come from the variation of the true value of the solar $\Delta m_{21}^2$ within the $3 \sigma$-allowed region as given in the figure caption, and the arrows correspond to the LMA best-fit values. This figure demonstrates that the exclusive $3 \, 000 \, \mathrm{km}$ option is only very good for CP violation, whereas the $\sin^2 2 \theta_{13}$ sensitivity strongly depends on the $\Delta m_{21}^2$-dependent ability to resolve the $(\delta,\theta_{13})$-degeneracy. The sign of $\Delta m_{31}^2$ can hardly be measured for large values of $\Delta m_{21}^2$ especially due to the $\mathrm{sgn}(\Delta m_{31}^2)$-degeneracy~\cite{Huber:2002mx}. The exclusive $7 \, 500 \, \mathrm{km}$ option performs very well for $\sin^2 2 \theta_{13}$ and the sign of $\Delta m_{31}^2$, since at the magic baseline the first term in \eq~(\ref{equ:PROBMATTER}) is measured in a clean way without being spoilt by degeneracies. However, it does not allow a measurement of $\delta_{\mathrm{CP}}$ because of the suppression of the CP-terms. The combination of the magic with the $3 \, 000 \, \mathrm{km}$-baseline turns out to allow, even for the worst case of $\Delta m_{21}^2$, very good sensitivities below $\sin^2 2 \theta_{13} \lesssim 10^{-4}$ for all of the quantities, because all degeneracies originating in $\delta_{\mathrm{CP}}$ can be resolved. Thus, taking all the information together, the $7 \, 500 \, \mathrm{km}+3 \, 000 \, \mathrm{km}$ option allows the best measurement of all of the investigated quantities without the planning risk of not knowing the exact values of the oscillation parameters.


In summary, we have demonstrated that choosing a specific neutrino factory baseline, which only depends on the matter density profile, allows a clean, almost degeneracy-free measurement of $\sin^2 2 \theta_{13}$ and $\mathrm{sgn}(\Delta m_{31}^2)$. We call this baseline the ``magic'' baseline, which is around $7 \, 300 \, \mathrm{km} - 7 \, 600 \, \mathrm{km}$, a distance, which corresponds, for example, to the baseline Fermilab--Gran Sasso. Since it is also important to have sensitivity to the CP phase, we find that the natural combination of two baselines at a neutrino factory favors the magic baseline combined with a $3 \, 000 \, \mathrm{km}$-baseline. Compared to superbeams, this option allows sensitivities to $\sin^2 2 \theta_{13}$, the sign of $\Delta m_{31}^2$, and CP violation even below $\sin^2 2 \theta_{13} \lesssim 10^{-4}$ ($3 \sigma$ confidence level) for a complete analysis including systematics, correlations, and degeneracies. Since the magic baseline is independent of the oscillation parameters, it leads to a very precise measurement of the relevant quantities independent of the actual values of the solar and atmospheric parameters. Thus, one can start looking for the neutrino factory and detector sites by already knowing at least one of the two baselines right now. 

We would like to thank M.~Lindner and T.~Schwetz for useful discussions and comments.
This work was supported by the ``Studienstiftung des deutschen Volkes'' [W.W.] and the ``Sonderforschungsbereich 375 f{\"u}r Astro-Teilchenphysik der Deutschen Forschungsgemeinschaft''.

\end{document}